# Robotic Telescope Labs for Survey-Level Undergraduates

Daniel E. Reichart

*University of North Carolina at Chapel Hill, Department of Physics and Astronomy, Campus Box 3255, Chapel Hill, NC 27599-3255*

**Abstract**. For the past dozen years, UNC-Chapel Hill has been developing a unique, survey-level astronomy curriculum, primarily for undergraduate students, with the goal of significantly boosting STEM enrollments on a national scale, as well as boosting students' technical and research skills. Called "Our Place In Space!", or OPIS!, this curriculum leverages "Skynet" – a global network of ≈2 dozen, fully automated, or robotic, professional-grade telescopes that we have deployed at some of the world's best observing sites. The curriculum has now been adopted by ≈2 dozen institutions, and we have just received $1.85M from NSF's IUSE program to expand it nationwide, with funding for participating instructors. The curriculum works equally well online as in person.

## 1. Skynet and PROMPT

Funded primarily by NSF, UNC-Chapel Hill began building "Skynet" and "PROMPT" in 2004. Skynet is sophisticated telescope control and queue scheduling software, that provides capacity to be a novel educational technology.[1] Skynet can control scores of telescopes simultaneously, allowing them to function individually or as an integrated whole. Furthermore, Skynet can control most commercially available telescope hardware, and provides participating institutions with easy-to-use web and API interfaces. Participating institutions are not charged, but instead contribute 10% of each of their telescopes' time for Director Discretionary science (e.g., gravitational-wave (GW) sources) and education (e.g., OPIS!; see below). The Skynet Robotic Telescope Network has grown to number ≈20 telescopes, with another ≈10 scheduled for integration over the next few years. Skynet optical telescopes range in size from 14" to 40" and span four continents and five countries. Skynet also includes a 20m-diameter radio telescope (see Figure 1) [1], with plans to integrate more.

PROMPT is a subset of the Skynet Robotic Telescope Network, consisting of our highest-quality telescopes at our highest-quality sites. Originally only at Cerro Tololo Inter-American Observatory in the Chilean Andes, PROMPT now spans three and soon five dark sites, in Chile, Australia (for near-continuous observing in the southern hemisphere), and Canada (for full-sky coverage; Figure 1).

---

[1] At https://skynet.unc.edu (password protected). For tutorials videos: https://tinyurl.com/skynet-tutorials.

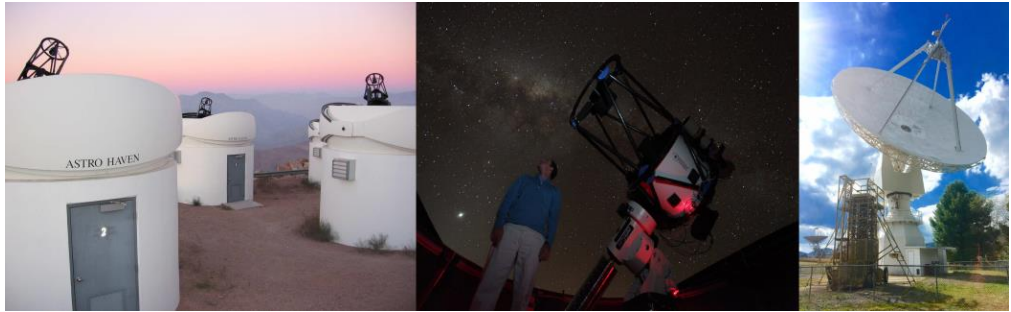

Figure 1. *Left: Three of the six original 16"-diameter Skynet/PROMPT telescopes at Cerro Tololo Inter-American Observatory (CTIO) in Chile. Center: One of the three upgraded 24"-diameter telescopes at CTIO (displaced originals have been relocated to other dark sites around the world). Right: Skynet's 20m-diameter radio telescope at Green Bank Observatory (GBO) in West Virginia.*

Skynet/PROMPT was originally built to carry out simultaneous, multi-wavelength observations of gamma-ray bursts (GRBs) when they are only tens of seconds old (to date, Skynet has observed 88 GRBs within 15 – 70 seconds (90% range) of spacecraft notification, detecting 50 optical afterglows on this timescale). However, our mission is now considerably broader. Skynet now serves state, national, and international user communities as a broad-based platform for small-telescope science: In addition to being used to study GRBs, Skynet, often in campaigns with other optical and radio telescopes around the world, and also with space telescopes, is being used to study GW sources, fast radio bursts, blazars, supernovae, supernova remnants, novae, pulsating white dwarfs and hot subdwarfs, a wide variety of variable stars, a wide variety of binary stars, exoplanetary systems, trans-Neptunian objects and Centaurs, asteroids, and near-Earth objects (Skynet is also the leading tracker of NEOs in the southern hemisphere; R. Holmes, private communication). Skynet data are now published in peer-reviewed journals every ≈20 days, including five times to date in Nature and Science. Furthermore, these publications have been of increasingly high impact, so far resulting in over 5,000 citations.

To date, roughly 40,000 students have used Skynet, with most now participating in three large, NSF-funded programs: (1) Skynet Junior Scholars (SJS), which is being carried out in partnership with 4H, for middle-school-age students; (2) Innovators Developing Accessible Tools for Astronomy (IDATA), for high-school students exploring computational thinking in astronomy and helping to develop image-analysis software for blind and visually impaired users (see §3, §7); and (3) Our Place In Space! (OPIS!), a Skynet-based laboratory curriculum for undergraduates in small to very large, introductory survey courses, which has now been adopted by ≈2 dozen institutions.

## 2. OPIS!

OPIS! is a Skynet-based laboratory curriculum for non-majors and potential majors alike. OPIS! consists of eight, and soon nine, labs in which students use the same research instrumentation as professionals to collect their own data. They then use this self-collected data (astronomical images and spectra) to reproduce some of the greatest astronomical discoveries of the past 400 years, and gain technical and research skills at the same time. Although students are not carrying out cutting-edge research, they are using cutting-edge research instrumentation, and consequently there is great overlap with the Course-based Undergraduate Research Experience (CURE) pathway model [e.g., 2]. And these labs/observing experiences are specifically designed to pair with standard introductory astronomy curricula, facilitating widespread adoption.

OPIS! is built around the cosmic distance ladder, which serves as an organizing principle in most introductory astronomy courses/sequences, and as such, it reinforces students' classroom experiences. The goal of OPIS! is to move beyond laboratory experiences in which students learn how to use a telescope for its own sake, to instead using them to do science, and furthermore, to do the same science that they are learning in class.

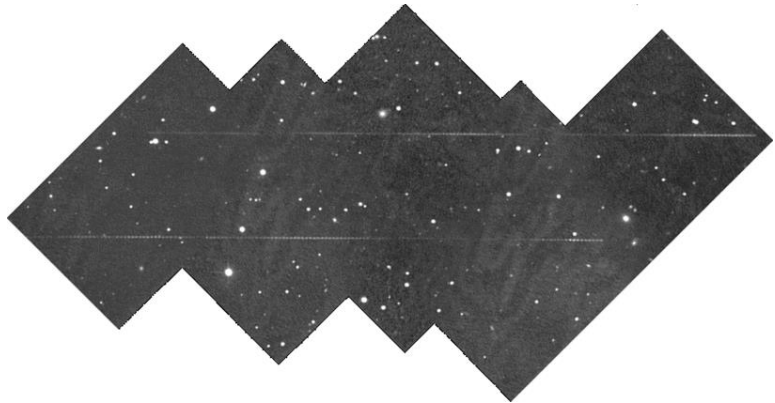

Figure 2. *Mosaic of images of NEO 2001 FE90 obtained simultaneously from PROMPT-CTIO in Chile and Appalachian State University's Dark Sky Observatory in North Carolina, which is part of Skynet. A parallactic shift of ≈8' and a rotational period of ≈30 minutes can be measured from the images.*

After an introductory lab in which students learn how to use (1) Skynet and (2) our image-analysis application, Afterglow (see §3), students, sometimes individually and sometimes collaboratively as a class, collect and measure their own images to distinguish between geocentric and heliocentric models using the phase and angular size of Venus (Lab 3), to measure the mass of a Jovian planet using the orbit of one of its moons and Kepler's Third Law (Lab 3), to measure the distance to an asteroid using parallax measured simultaneously by Skynet telescopes in different hemispheres (Lab 4; e.g., see Figure 2), and to measure the distance to a globular cluster using an RR Lyrae star as a standard candle (Lab 5). More is done with archival data that takes longer than a semester to collect (e.g., Cepheid stars, Type Ia supernovae, etc.). See Figure 3 for more.

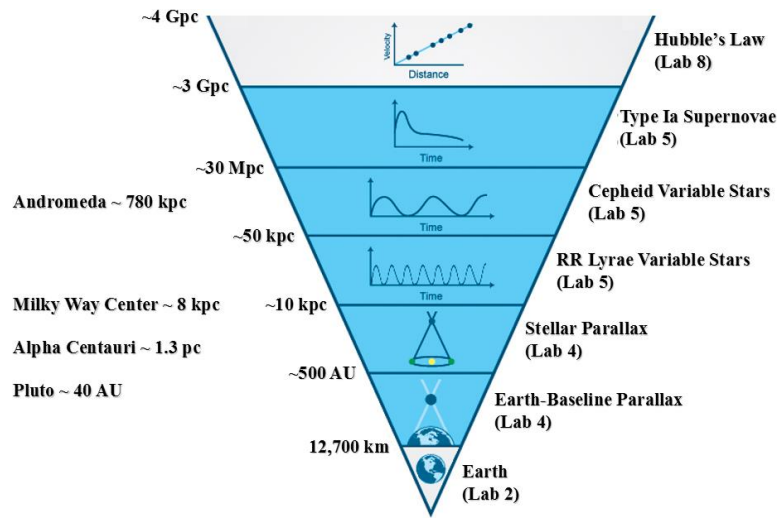

Figure 3.  *OPIS! is a series of eight, interconnected labs that teach the evolution of our understanding of Earth's place in the universe, and the cosmic distance ladder. Lab 6 makes use of what was learned in Lab 5 to teach the Great Debate.  In Lab 7, students use Skynet to collect 21-cm spectroscopy from GBO's 20m-diameter radio telescope (Figure 1) to measure the Galaxy's rotation curve and mass distribution [e.g., 3].  A ninth, capstone experience on multi-messenger astronomy is being developed, in which students use Skynet to collect comparison images to identify the optical counterpart to a GW event, and then use the GW event as a standard siren, and a provided spectrum of the counterpart, to measure Hubble's Constant (Skynet was the second of six groups to co-discover the first optical counterpart to a GW event [4-6].)*

The introduction of this curriculum at UNC-Chapel Hill twelve years ago resulted in a >100% increase in enrollment over a five-year period – now one in six UNC-Chapel Hill undergraduates take at least one of our introductory astronomy courses.  It additionally contributed to an ≈300% increase in astronomy-track majors and minors (from ≈10 to ≈40, ≈50% female), and to our department being awarded two new tenure-track astronomy hires.

OPIS! has since been adopted by ≈2 dozen institutions, including large R1 institutions, smaller undergraduate-only institutions, minority-serving institutions, rural-serving institutions, community colleges, and advanced, college-preparatory high schools.  And it is used in a variety of formats, including integrated into the classroom, as a separate/stand-alone laboratory experience, and fully online (see §6).

Funded by an NSF TUES award, OPIS! was the subject of a preliminary/exploratory study across ten of these institutions, and across all of these formats.  The study found that once obvious factors, such as the grade that each student expected to receive, and their career plans, were controlled for, Skynet-based observing experiences were one of only two course components that led to a statistically significant improvement in STEM attitudes [7].  For example, traditional telescope labs, and non-telescope labs, as well as in-class activities known to yield learning gains, did not have a similar effect on attitudes.  Funded by our NSF IUSE award, a larger, Type IV case study, addressing not only students' attitudes, but self-efficacy and conceptual knowledge in astronomy and STEM, and encompassing three times as many institutions, is now underway.

Finally, OPIS! is designed to scale: Currently, OPIS! serves ≈2,000 students/yr, but we could serve tens of thousands per year, or more, without compromising professional operations.

## 3. Afterglow Access

Skynet allows students to acquire professional-quality images from multiple, professional-quality telescopes and sites around the world. However, this is only half the battle. Students also need to be able to explore their images, and make fundamental measurements from them, around which relevant laboratory experiences can be designed. As such, we have additionally developed *Afterglow Access*.[2]

*Afterglow Access* is a web application, written in AngularJS. The advantage of being a web application is that students do not need to install it, and updates can be done server-side. Furthermore, *Afterglow Access* is connected to Skynet's 100 TB RAID, so students do not need to download, independently store, and re-upload their images. Nor do they even need a quality computer, as the heaviest computational lifting (e.g., processing/analyzing many images simultaneously) is handled server-side as well. Doing it ourselves also ensured that all of the capabilities that we needed for OPIS! (and IDATA; §1) would be available in a single package, and that this package would be at an appropriate, student level.

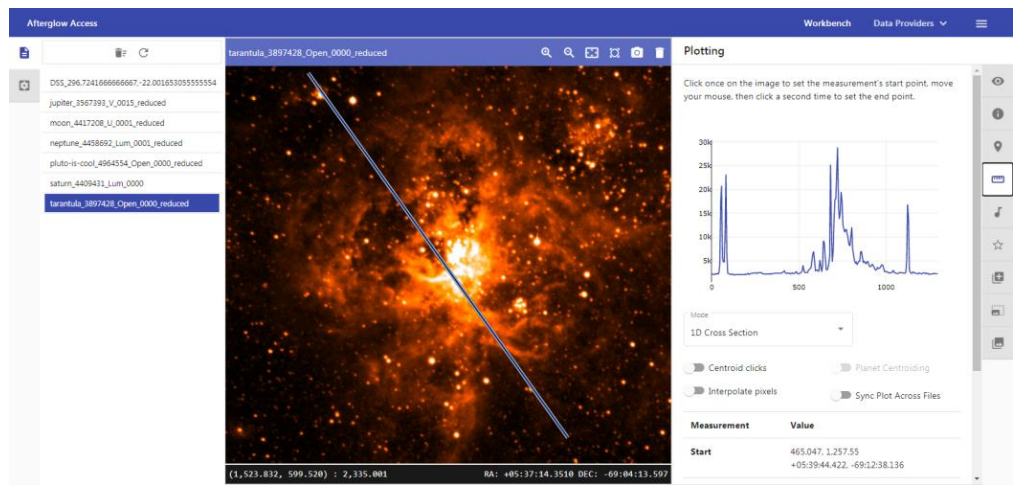

Figure 4. *Screenshot of Afterglow Access's angular distance measurement tool.*

*Afterglow Access*'s capabilities currently include everything students need to carry out the above curricula, including: (1) image exploration (e.g., bright-ness/contrast rescaling to bring out fainter objects/details, marking and labeling tools, etc.); (2) measuring angular distances, both across and between astronomical objects (see Figure 4); (3) image alignment and arithmetic (e.g., averaging images, to identify moving objects, or differencing images, to detect brightening/fading objects); (4) photometry (both of multiple objects and across multiple images); and (5) plotting light curves and calculating/plotting periodograms and period-folded light curves (e.g., to study rotating asteroids, variable stars, etc.).

---

[2] At https://afterglow.skynet.unc.edu (password protected). For a demonstration video: https://tinyurl.com/afterglow-demo. For tutorial videos: https://tinyurl.com/afterglow-tutorials

New capabilities currently in development include the ability to upload non-Skynet images; direct access to most major archives; automated, professional-quality photometric calibration of images, using these same archives; RGB/LRGB color combination; and radio data and image processing.

## 4.   Custom Graphing Application

OPIS! comes with its own, custom graphing utility, which meets all of the curriculum's graphing needs.[3] This includes options for plotting multiple curves, both with and without common x values; for determining whether the solar system is geocentric or heliocentric, from measurements of Venus's angular diameter and phase; for chi-by-eye fitting of orbits, to extract semi-major axes and orbital periods of moons (see Figure 5); for plotting and period-folding light curves of variable stars and rotating asteroids; for calculating and producing projected, top-down maps of the Milky Way's globular cluster distribution; and for plotting HI (21-cm) radio spectra and measuring redshifts.  (The labs also teach students some basic spreadsheet skills.)

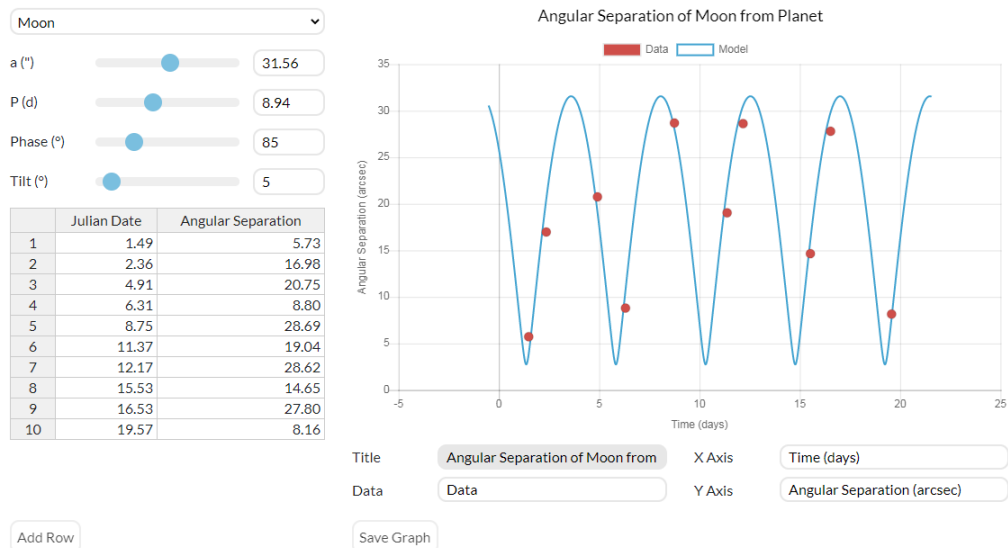

Figure 5.  *Screenshot of the OPIS! graphing utility's orbit-fitting application.*

---

[3] At https://tinyurl.com/opis-graph.   For tutorial videos:   https://tinyurl.com/graphing-tutorials.

## 5. Overview and Tutorial Videos

A recent addition to the curriculum is ≈65 overview and tutorial videos, which are embedded in the labs, but are also available here: https://tinyurl.com/opis-videos. They are currently garnering ≈400 views per day and ≈30 hours viewed per day on YouTube. They can be used in multiple ways. For some instructors, they are helpful supplements that students can consult when completing labs. Others use them as the primary mode of lab instruction, freeing themselves up to focus more, and more time, on lecture/regular instruction, and on working with students one-on-one as they need it.

We are currently producing "enhanced" versions of these videos, which incorporate zooms, pointers, reserved space for closed captioning, etc., to increase accessibility for deaf and hard-of-hearing (DHH) and visually impaired students alike (see §7).

## 6. Formats, Auto-grading Support, and Cost/Sustainability

OPIS! works equally well online as in person. This is because Skynet and *Afterglow Access* are already fully online/distance-learning tools. Indeed, UNC-Chapel Hill has offered both in-person and online sections of OPIS! every year for the past decade. And when the pandemic struck in Spring 2020, OPIS! sections, at ≈2 dozen institutions, were among the few science lab courses at these institutions that were able to transition to fully online without difficulty.

Given this, we have partnered with homework-assessment company Cengage/WebAssign, which has packaged the OPIS! labs into an interactive, and maximally auto-graded,[4] online manual/text. Most participating institutions choose this product, both for its ease of use, and because it greatly simplifies grading, particularly large numbers of students. It has an ISBN number and can be ordered by campus bookstores the same as any textbook (allowing students to use financial aid money instead of personal funds), and it now integrates into standard learning management systems (LMSs).

Cengage/WebAssign charges for this product (typically $65/student), but a portion of this cost is returned to UNC-Chapel Hill to support telescope operations and maintenance, software maintenance and further development (see §8), etc. This gives OPIS! a long-term sustainability model that has allowed it to survive grant funding cycles. (Currently, Skynet costs ≈$350K/year to operate.) Simultaneously, it gives students unique access to research-grade equipment at some of the best and darkest sites around the world. (Telescope time can also be purchased separately, directly from UNC-Chapel Hill, in 10-hour blocks.)

---

[4] Short answers, and submitted images and graphs must still be graded manually, but numerical responses and calculations are evaluated automatically.

## 7. Accessibility

As part of our previous SJS and IDATA efforts (§1), we partnered with accessibility specialists at Yerkes Observatory, now employed by Geneva Lakes Astrophysics and STEAM (GLAS), to ensure that both (1) these awards' middle- and high-school curricula and (2) Skynet's supporting technologies are accessible to blind and visually impaired (BVI) and deaf and hard-of-hearing (DHH) students. One byproduct of these efforts is that both Skynet and *Afterglow Access* are already compatible with standard screen readers used by BVI students. Another is that we have developed and integrated into *Afterglow Access* a new, significant image-sonification capability. This uses frequency (and stereo projection, if available) to convey left-right information, time to convey up-down information, and volume to convey (monochromatic) intensity information, and, through GLAS, is already being used by BVI students to identify astronomical objects (e.g., star clusters vs. nebulae vs. galaxies) as well as image characteristics and artifacts (e.g., relative focus quality, a satellite streaking across the image, etc.). We also developed a screen reader-compatible interface that allows BVI students to zoom in to audio-distinct regions and re-sonify them, permitting image exploration. Future efforts will include the incorporation of timbre/tone for color images (e.g., violins for bluer regions and cellos for redder regions, though we will experiment with many options).

## 8. Instructor Training and Funding

Each fall, we offer weekly Zoom training sessions for adopting instructors, where we review one lab per week. We also record these sessions for instructors with time conflicts, and for instructors adopting off-semester.[5]

Each spring, we offer weekly Zoom sessions for instructors who have already implemented OPIS! at their institution. Here, we review updates to the curriculum and to the supporting software, and we collect detailed feedback so we can continue to improve the labs.

Funded by our NSF IUSE award, we have up to $3,500 for at least 35 adopting instructors to learn and implement the curriculum at their institution. This includes attending one semester of training sessions, one semester of feedback sessions, and administering two pre/post surveys to your students over the course of the 5-year award. One survey measures student attitudes, self-efficacy, and career intentions. The other measures content knowledge. The end result will likely be the largest study of the positive impacts of robotic-telescope use in education ever conducted, at least at the undergraduate level.

For additional information about OPIS! and implementing it at your institution: (1) https://tinyurl.com/opis-workshop, (2) https://tinyurl.com/opis-links, and/or (3) email introastro@unc.edu.

---

[5] Fall 2020's training sessions are here: https://tinyurl.com/opis-training

**Acknowledgements**. We gratefully acknowledge the support of the National Science Foundation, through the following programs and awards: ISE 1223235, HBCU-UP 1238809, TUES 1245383, STEM+C 1640131, AAG 2007853, IUSE 2013300. We are also appreciative to have been supported by the Mt. Cuba Astronomical Foundation, the North Carolina Space Grant Consortium, and the University of North Carolina System.